\documentclass{article}
\usepackage[T1]{fontenc} 
\usepackage[utf8]{inputenc} 
\usepackage{amsmath,cite,url}
\usepackage{graphicx}
\usepackage[lbd]{ismir}
\usepackage{kotex}
\usepackage{listings}
\usepackage{dblfloatfix}
\usepackage{color}

\usepackage[firstpage]{draftwatermark}
\definecolor{lightgray}{rgb}{0.9,0.9,0.9}
\definecolor{darkgray}{rgb}{0.4,0.4,0.4}
\SetWatermarkFontSize{12pt}
\SetWatermarkScale{1.1}
\SetWatermarkAngle{90}
\SetWatermarkHorCenter{202mm}
\SetWatermarkVerCenter{170mm}
\SetWatermarkColor{darkgray}
\SetWatermarkText{Late-Breaking / Demo Session Extended Abstract, ISMIR 2025 Conference}

\title{WaveRoll: JavaScript Library for Comparative MIDI Piano-Roll Visualization}

\twoauthors
 {Hannah Park} {Sogang University \\ Department of Artificial Intelligence \\ {\tt{hannah@sogang.ac.kr}}}
 {Dasaem Jeong} {Sogang University \\ Department of Art \& Technology \\ {\tt{dasaemj@sogang.ac.kr
}}}

\def\authorname{H. Park, and D. Jeong}

\usepackage[pdfauthor={\authorname},hidelinks,bookmarks=false]{hyperref}

\begin{document}

\maketitle

\begin{abstract}

\textbf{WaveRoll} is an interactive JavaScript library that enables comparative visualization and synchronized playback of multiple MIDI piano rolls on a browser. 
It addresses a specific evaluation need in Automatic Music Transcription (AMT), contrasting multiple MIDI outputs produced from the same input. 
The library displays multiple MIDI tracks on a single, time-aligned grid with synchronized audio, allowing users to compare pitch and timing, identify missed or extra notes, and observe onset and offset differences, as well as section-level patterns.
We expect that such comparisons would assist in model evaluation and error analysis, and help readers to understand the model behavior better. 
The open-source library is available at \url{https://github.com/crescent-stdio/wave-roll}

\end{abstract}
\section{Motivation \& Related Works}\label{sec:introduction}

In Music Information Retrieval (MIR), Automatic Music Transcription (AMT) involves converting music audio into symbolic music data, such as MIDI. While quantitative metrics, such as precision, recall, and F1-score, are used to assess the performance of deep learning-based AMT models \cite{jamshidi2024AMTsurvey}, they do not fully reveal the nature of errors, including which specific notes are incorrect or whether errors occur consistently in certain sections. Therefore, to thoroughly understand a model's behavior, it is essential to visually and aurally examine the resulting MIDI data.

Current research on AMT primarily presents qualitative results through static or non-interactive media. For instance, some studies display input audio and resulting MIDI as a single video \cite{yan2024scoringTranskun}, while others provide the model's output for a Ground Truth as a piano roll image with an individual MIDI player \cite{cheuk2021reconvat}, or allow playback of a MIDI output for a Ground Truth audio \cite{onsetsAndFrames}.
However, these approaches have limitations. They make it difficult for users to compare outputs from multiple models for the same input simultaneously or to perform interactive analysis by seeking specific positions within the audio.

WaveRoll is an interactive JavaScript library that reconfigures the piano-roll style of music data visualization for multi-MIDI comparison on the web. It is proposed as a solution based on existing concepts: web-based MIDI piano-roll players (from \texttt{html-midi-player} \cite{cifkaohtmlmidiplayer}) and multi-track control (from \texttt{trackswitch.js} \cite{werner2017trackswitchjs}). This feature allows users to toggle, play, stop, and seek multiple MIDI files simultaneously on a single screen, adopting a style commonly found in Digital Audio Workstations (DAWs).

Users can concurrently load MIDI results generated by different models for the same audio input, providing an immediate means to identify discrepancies in pitch, timing, onset, and sustain. WaveRoll is delivered as a \texttt{<wave-roll>} custom web component, facilitating easy analysis of model performance in local development environments using Node Package Manager(npm)\footnote{\url{https://www.npmjs.com/package/wave-roll}} or via the library's demo page\footnote{\url{https://crescent-stdio.github.io/wave-roll}}. Furthermore, researchers can utilize WaveRoll to create web-based interactive demonstrations with minimal code, enabling readers to grasp their findings intuitively.

\section{WaveRoll}

\begin{figure*}[t]
 \centerline{
 \includegraphics[alt={Screenshots of WaveRoll components},width=1.0\textwidth]{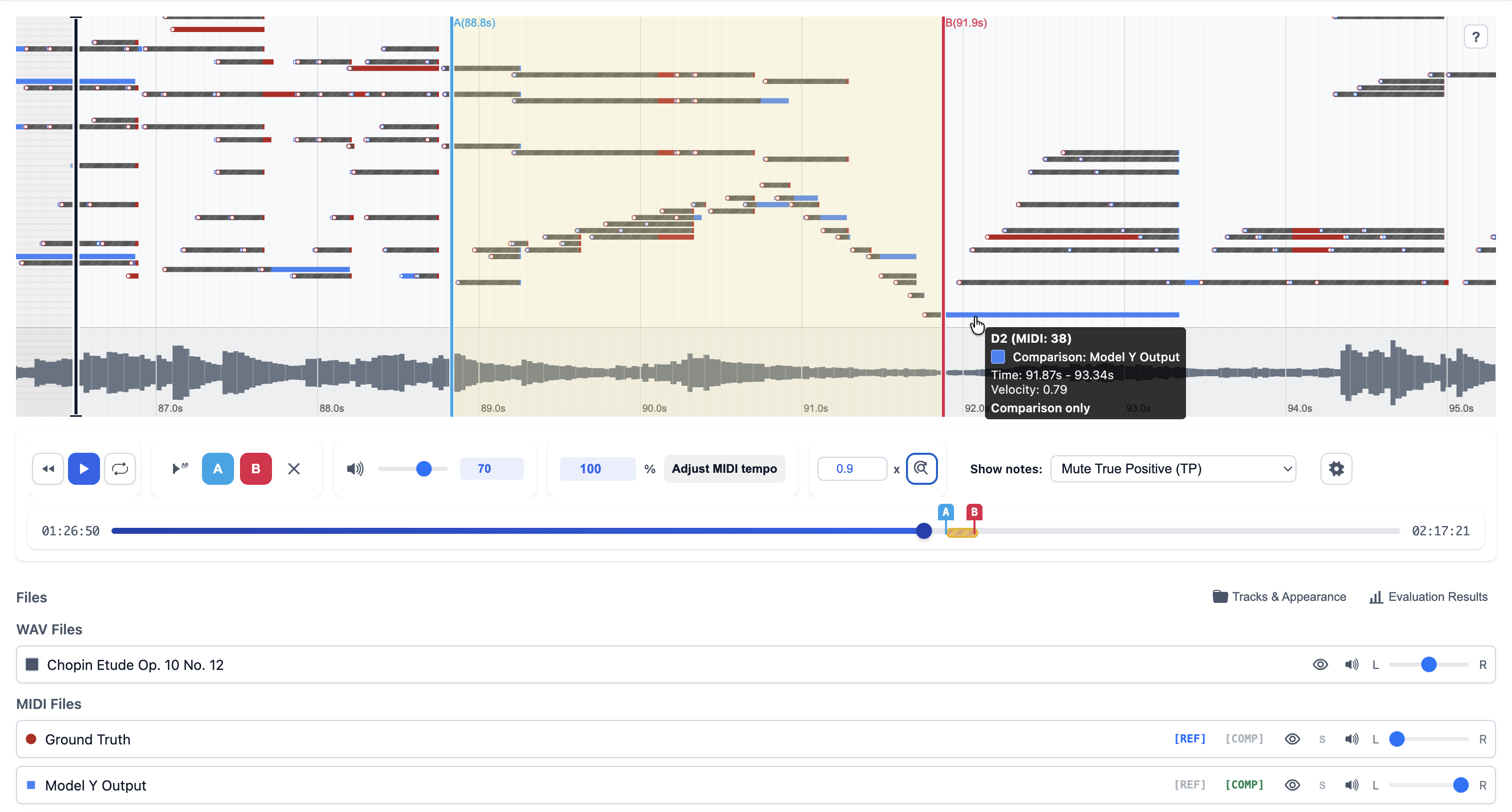}}
 \caption{The WaveRoll interface, featuring a layered piano roll and a side panel for audio controls.}
 \label{fig:wave-roll}
\end{figure*}

\begin{figure}[!t]
\begin{lstlisting}[
    language=HTML,
    basicstyle=\footnotesize\ttfamily, % Smaller font size
    breaklines=true,                 % Automatic line breaking
    breakatwhitespace=false,         % Allow breaking within strings (for long URLs)
    frame=tb,                        % Add lines at the top and bottom
    captionpos=b,                    % Position caption at the bottom
    showstringspaces=false,          % Do not show spaces in strings
    tabsize=2,                       % Set tab size to 2
    keepspaces=true,
]
<!-- Add the library -->
<script type="module">
  import 'https://cdn.jsdelivr.net/npm/wave-roll@latest/dist/wave-roll.es.js';
</script>

<!-- Use the WaveRoll component -->
<wave-roll files='[
  {"path": "gt.wav", "name": "Audio",
   "type": "audio"},
  {"path": "gt.mid", "name": "Ground Truth",
   "type": "midi"},
  {"path": "my_model.mid", "name": "My Model",
   "type": "midi"}
]'></wave-roll>
\end{lstlisting}
\label{lst:waveroll}
\caption{Example of WaveRoll implementation in HTML.}
\end{figure}

WaveRoll introduces the \textit{Layered Piano Roll} uniquely displays multiple MIDI files as distinct visual layers on a single, unified grid as illustrated in Figure \ref{fig:wave-roll}.

\subsection{Key Insight Features}
WaveRoll is designed to help users gain comprehensive insights by integrating visual, auditory, and note-level analysis.

\subsubsection{Interactive Visual Analysis}
The library provides an interactive environment for visual analysis. Users can view multiple MIDI tracks on a layered canvas, controlling the visibility of each track's piano roll and sustain data while using zoom and pan for detailed examination. To aid comparison, each track can be assigned a custom color. Furthermore, various highlighting modes can visually emphasize matching or differing notes between tracks, allowing for a more focused analysis of similarities and discrepancies in model outputs.

\subsubsection{Detailed Auditory Comparison}
The library integrates not only the visualization but also the playback of original audio waveforms alongside MIDI data.
For in-depth auditory analysis, it offers Mute and Panning controls for individual tracks. Crucially, the A-B Looping feature enables users to repeat specific sections, allowing for a focused comparison of subtle rhythmic or harmonic variations that may be challenging to identify through visual inspection alone.

\subsubsection{Visual Note Matching}
WaveRoll facilitates qualitative judgment through note-level comparison. Based on the note evaluation method from \texttt{mir\_eval}\cite{Raffel2014mireval}, the library offers a visual highlighting feature for matching notes between two tracks, configurable by a user-defined tolerance. This visualization is beneficial for analyzing model accuracy and identifying error patterns in Automatic Music Transcription.

\subsection{Implementation}

Using WaveRoll on a webpage is straightforward. It is available as an npm package, installed with the command \texttt{npm install wave-roll}. As illustrated in Figure \ref{lst:waveroll}, developers can integrate the component into their HTML and provide the MIDI files for comparison as a JSON array to the \texttt{files} attribute. 
It can also be embedded in Jupyter notebooks using \texttt{IFrame}.

\section{Conclusion}
WaveRoll is designed to enhance the qualitative evaluation of AMT research. It addresses current limitations by visualizing multiple MIDI files on a layered piano roll with synchronized playback. This functionality allows researchers to intuitively analyze model error patterns and helps readers more easily understand the results.

Future developments for WaveRoll include improved user experience and custom SoundFonts.
While its primary focus is currently AMT evaluation, WaveRoll's application can be extended to other research areas that require the comparison of multiple outputs, such as Symbolic Music Generation. As an open-source project, community contributions are highly encouraged.

\section{Acknowledgement}
This work was supported by the Ministry of Education of the Republic of Korea and the National Research Foundation of Korea (NRF-2024S1A5C3A03046168).

\bibliography{ISMIRtemplate}

\end{document}